\documentclass[twocolumn,nopacs,english,prl,superscriptaddress,preprintnumbers,amsmath,amssymb,floatfix]{revtex4}

\usepackage{graphicx,epsf}
\usepackage{color}

\makeatletter
\def\NAT@bibsetnum#1{%
 \setlength{\topsep}{\z@}%
 \NATx@bibsetnum{#1}%
}%
\renewenvironment{thebibliography}[1]{%
 \NAT@thebibliography{#1}%
 \@clubpenalty\clubpenalty
 \let\@TBN@opr\present@bibnote
 \@FMN@list
}{%
 \@endnotesinbib
 \edef\@currentlabel{\arabic{NAT@ctr}}%
 \NAT@endthebibliography
 \global\let\auto@bib\@empty   
}
\newcommand*{\supplementarystart}{%
  \close@column@grid%
  \clearpage%
  \onecolumngrid%
  \setcounter{enumiv}{0} % resets counter for references
  \setcounter{equation}{0} % resets counter for equations
  \setcounter{figure}{0} % resets counter for figs
  \setcounter{table}{0} % resets counter for tables
  \setcounter{page}{1}
  \c@secnumdepth=4
  \renewcommand{\theequation}{S\arabic{equation}} % equations numbered with S...
  \renewcommand{\bibnumfmt}[1]{[s##1]} % bibtems [S...]
  \renewcommand{\@onlinecite}{s\citealp} % citations [S...]
  \renewcommand{\cite}[1]{{[}\onlinecite{##1}{]}}
  \renewcommand{\thefigure}{s\arabic{figure}}
  \renewcommand{\thetable}{s\Roman{table}}
  \renewcommand{\thepage}{s\arabic{page}}
}
\makeatother

\begin{document}

\title{
Sinai Diffusion at Quasi-1D Topological Phase Transitions\\
}

\author{Dmitry Bagrets}

\affiliation{Institut f\"ur Theoretische Physik, Universit\"at zu K\"oln,
Z\"ulpicher Stra\ss e 77, 50937 K\"oln, Germany}

\author{Alexander Altland}

\affiliation{Institut f\"ur Theoretische Physik, Universit\"at zu K\"oln,
Z\"ulpicher Stra\ss e 77, 50937 K\"oln, Germany}

\author{Alex Kamenev}

\affiliation{W. I. Fine Theoretical Physics Institute and School of Physics and Astronomy, University
of Minnesota, Minneapolis, MN 55455, USA}

\date{\today}

\begin{abstract}
We consider critical quantum transport in disordered topological quantum wires at the transition between phases with different topological indexes. Focusing on the example of thermal transport in class D (`Majorana') quantum wires, we identify a transport universality class distinguished for anomalous retardation in the propagation of excitations -- a quantum generalization of Sinai diffusion. We discuss the expected manifestations of this transport mechanism for heat propagation in topological superconductors near criticality and provide a microscopic theory explaining the phenomenon. 
\end{abstract}

%\pacs{}

\maketitle

%%%%%%%%%%%%%%%%%%%%%%%%%%%%%%%%%%%%%%%%%%%%%%%%%%%%%%%%%%%%%%%%%%%%%%%

\begin{figure}[b]
\includegraphics[width=7.0cm]{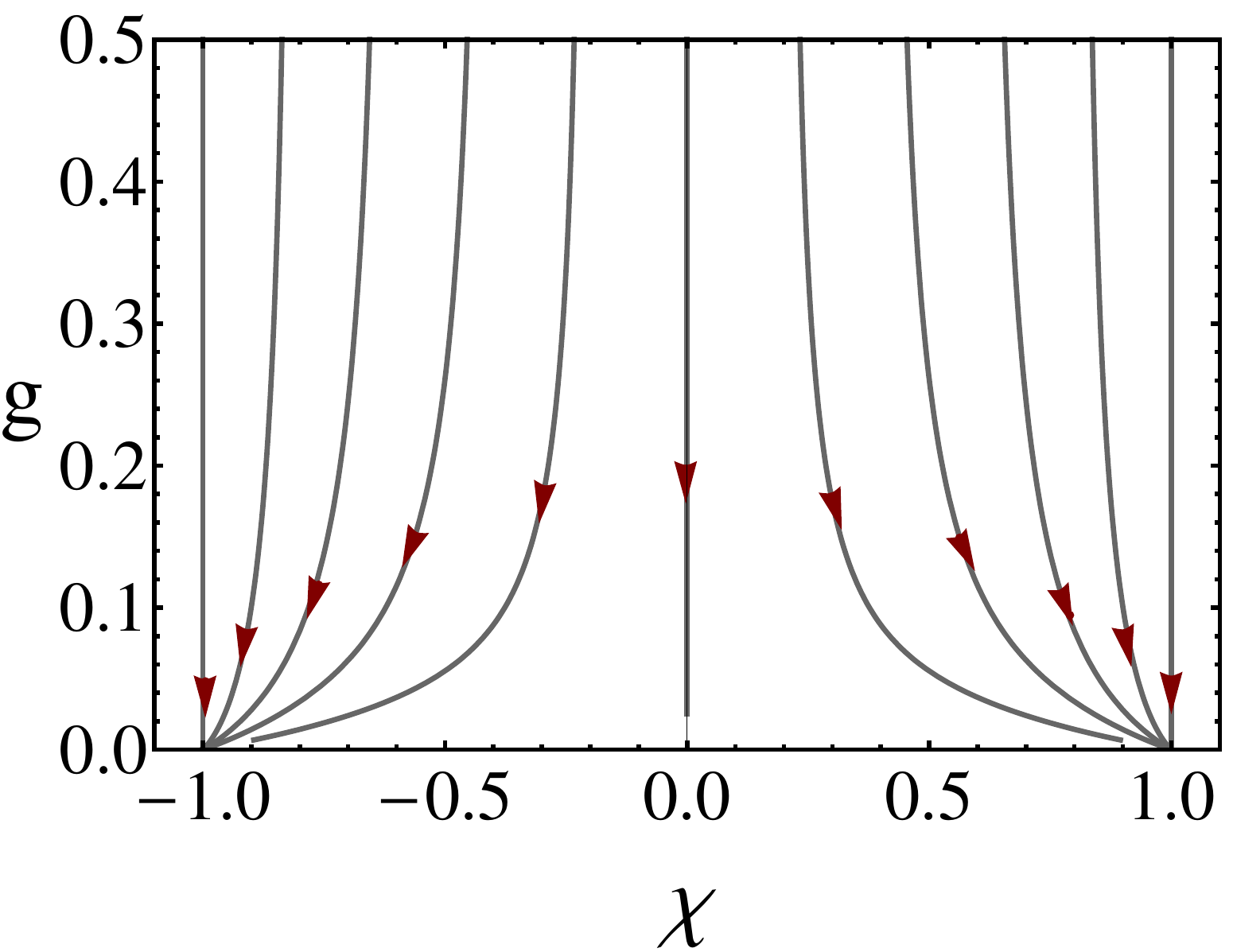}
\caption{Flow of the conductance $g$ and the average $\Bbb{Z}_2$ topological index $\chi$ (the kink's fugacity) 
as a function of system size for class D system. 
}
\label{fig:flow}
\end{figure}

Topological insulators  and superconductors (TI) are a novel form of quantum matter
distinguished by the presence of topological twists in their band structure~\cite{Schnyder:2008,Kitaev:2009}. 
While many of the salient features of these materials have first been
understood  within the idealized framework of 
clean models~\cite{Hasan:2010,Qi:2011,Teo:2010,Stone:2011,Kennedy:2015}, more recently the
disordered topological insulator has become a focus of 
attention~\cite{Groth:2009,Shindou:2009,Ryu:2012,Prodan:2014,Kobayashi:2014,Beenakker:2015,Morimoto:2015}. 

Besides the fact that every realistic system
is disordered, one of the driving forces behind this generalization is \emph{universality}. 
Indeed, upon averaging over a
distribution of impurities, system specific details become inessential and the
physics of TIs is reduced to its core: an interplay of symmetries and topology~\cite{Mirlin:2010}. 
This universality paradigm finds its ultimate expression in a two-parameter scaling
picture, which describes both the conduction properties of the topological phases and 
the expected values of their topological indices in terms of the flow diagram, exemplified 
in Fig.~\ref{fig:flow} for the $\Bbb{Z}_2$ superconducting class D.
First introduced within the context of the integer quantum Hall system~\cite{Khmelnitskii:83,Levine:1984} -- nowadays
categorized as a 2D class $\mathrm{A}$ topological insulator  -- the two
parameter scheme describes the competition between topology and localization. 
Most notably in a critical regime near a phase transition between two distinct topological 
phases, Anderson localization, otherwise dominant in low-dimensional systems, is overpowered by topological correlations, giving way to the critical delocalized state. 
Specifically, for the five symmetry classes of \emph{quasi-1D quantum
wires} it has been possible to map out the ensuing two-parameter flow diagrams by
non-perturbative field theoretical calculations~\cite{Altland:2015}.

In this Letter we address the question what kind of \emph{transport mechanisms} govern low-dimensional topological insulators in the critical regime and how they manifest themselves in concretely described  (thermal) conduction probes. 
Focusing on the quasi-1D case, we  show that the compromise between a general tendency to localize and the topologically
enforced system-wide formation of correlations at criticality results in the 
ultra-retarded transport mechanism -- a quantum extension of classical \emph{Sinai diffusion}. The latter is  distinguished by anomalously slow spreading of displacement $x$  in time $t$ as 
\begin{equation}
\label{eq:Sinai}
x\propto \xi_0\log^2 t\,,
\end{equation}
instead of the conventional $x\propto \sqrt{Dt}$, where $\xi_0$ is a correlation length and $D$ the diffusion constant. 
{We also explore how this critical transport  connects to the strongly Anderson localized
off-critical regimes}, where  correlations
are due to resonant transitions between remote localized states. The overall result
is a comprehensive picture of dynamical correlations in  topological quasi-1D Anderson insulators. We argue that essential elements of this picture are accessible via established procedures of 
`thermal electronics'~\cite{Giazotto:2006}, specifically the tunnel probe detection 
of a non-equilibrium electron distributions combined with the time-resolved measurements.

Sinai diffusion was introduced~\cite{Sinai:1982,Bouchaud:1990,Comtet:1998} within the context
of thermal Langevin dynamics in the presence of a quenched random force. First evidence
that this classical transport mechanism  might be of relevance to 
quantum models follows from pioneering earlier work on one-dimensional random mass
Dirac fermions~\cite{Balents:1997}, or equivalent models of random Ising chains~\cite{Shankar:1987}. 
Consider a random mass Dirac equation
\begin{equation}
\label{eq:Dirac}
E\Psi = \Big[-i\sigma_2\partial_x + \sigma_1 m(x)\Big]\Psi, 
\end{equation}
where $\Psi=(\psi_+,\psi_-)^T$ is a two component spinor, $\langle m(x)\rangle=\bar
m$, and $\langle m(x)m(x') \rangle = 2\xi_0^{-1}\delta(x-x')$. For $\bar m\not=0$ the Hamiltonian entering the equation defines a toy model of a single channel class $\mathrm{BDI}$ topological quantum wire with the index $\mathrm{sgn}(\bar m)$. For $\bar m\to 0$ the model becomes critical and gapless, but at the same time the absence of a large energy scale (`Fermi energy') means that any amount of randomness, $\xi_0<\infty$, has a strong effect on transport. Ref.~\cite{Balents:1997} confirmed this anticipation by applying supersymmetry techniques to compute a diverging dynamical exponent $z\to \infty$ at $E=0$, meaning infinitely slow transport. A heuristic way to reach this conclusion is to observe that in the band center,  $E=0$, the equation is solved by   $\psi_\pm(x)\propto
\exp\{\mp \int^x m(x) dx\}$. The $+(-)$-component of these real (Majorana) solutions has its maximum $\partial_x\psi_\pm(x)=0$ at a point $x_0$ where 
$m(x_0)=0$ and $m'(x_0)<0$ $(m'(x_0)>0)$. Away from these points, a typical critical wave function 
($\bar m=0$) decays as $\psi_\pm(x)\propto
\exp\{-\sqrt{|x-x_0|/\xi_0}\}$. For  small deviations $\omega$ off $E=0$ this
qualitative behavior does not change and the corresponding system of low energy
states is expected to generate dynamical correlations  as described by a modified
Mott~\cite{Mott:1968,Mott:1970} argument:  propagation of dynamical excitations with 
frequency $\omega$  necessitates the presence of states separated in
energy by $\omega$. On the other hand the above construction implies that states
separated in space by $x$ overlap (repel each other in energy) as $\omega \sim
\exp\{-\sqrt{|x|/\xi_0}\}$. This leads to a correlation between spatial and
frequency scales, $x\sim \xi_0 \ln^2(1/\omega) \sim \xi_0 \ln^2 t$,
characteristic for Sinai diffusion. We finally point out that a formal link to the 
classical random force Langevin dynamics may be
drawn by mapping the corresponding  Fokker-Planck equation to the square 
of the effective random mass Dirac operator \cite{Bouchaud:1990}.

These constructions draw a connection between Sinai diffusion and the single random mass
Dirac quantum channel. However, it would be premature to jump to the conclusion that
the phenomenon survives under the more general conditions of a three-dimensional yet
spatially anisotropic \emph{quasi}-1D TI. In quasi-1D systems spatial
inhomogeneities are comparatively weaker as they can be circumnavigated  via local
three dimensional diffusion. Formally, this reflects in bare dynamical
exponents $z=2$ of the field theories describing  quasi-1D disordered conductors, 
suggesting that at a critical point a re-entrance to ordinary diffusive spreading of
excitations (which is not an option in the strictly 1D context) might take
place. The main finding of this work is that, contrary to this naive
expectation, it is Sinai diffusion that governs the critical points both
$\Bbb{Z}$ and $\Bbb{Z}_2$  {\em multi-channel} topological quantum wires in all five topologically non-trivial symmetry classes. The ensuing
$z=\infty$ dynamical correlations can be understood as limits of conventional
Anderson localized Mott correlations off criticality.

Before exploring the microscopic foundations of this result let us discuss how it
manifests itself in the critical transport properties of the TI quantum wire.
The discussion above suggests the opening of a  transport channel supported by the
resonant coupling of Majorana states forming at the band center of a  disordered  
system at criticality. The logarithmic dependence of the resonant coupling range on time (or inverse
frequency) indicates that two principles compete  in the contribution of this
channel to transport. On the one hand it takes an exponentially large time $t\sim
\exp(\sqrt{|x|/\xi_0})$ for resonant excitations to propagate into the system. On the
other hand, at large times, only a diminishing number of states $\propto \omega
\propto t^{-1}$ around the band center maintains the phase coherence necessary to
contribute to the transport. This suggests that the resonant channel 
contributes to quantum transport in a temporally non-monotonous way, where peak
levels are reached at exceptionally large times before the phase decoherence leads 
to the signal suppression. In the following we propose a
concrete setup designed to probe such type of behavior and make quantitative
predictions for transport near criticality.

\begin{figure}[t]
\includegraphics[width=7.0cm]{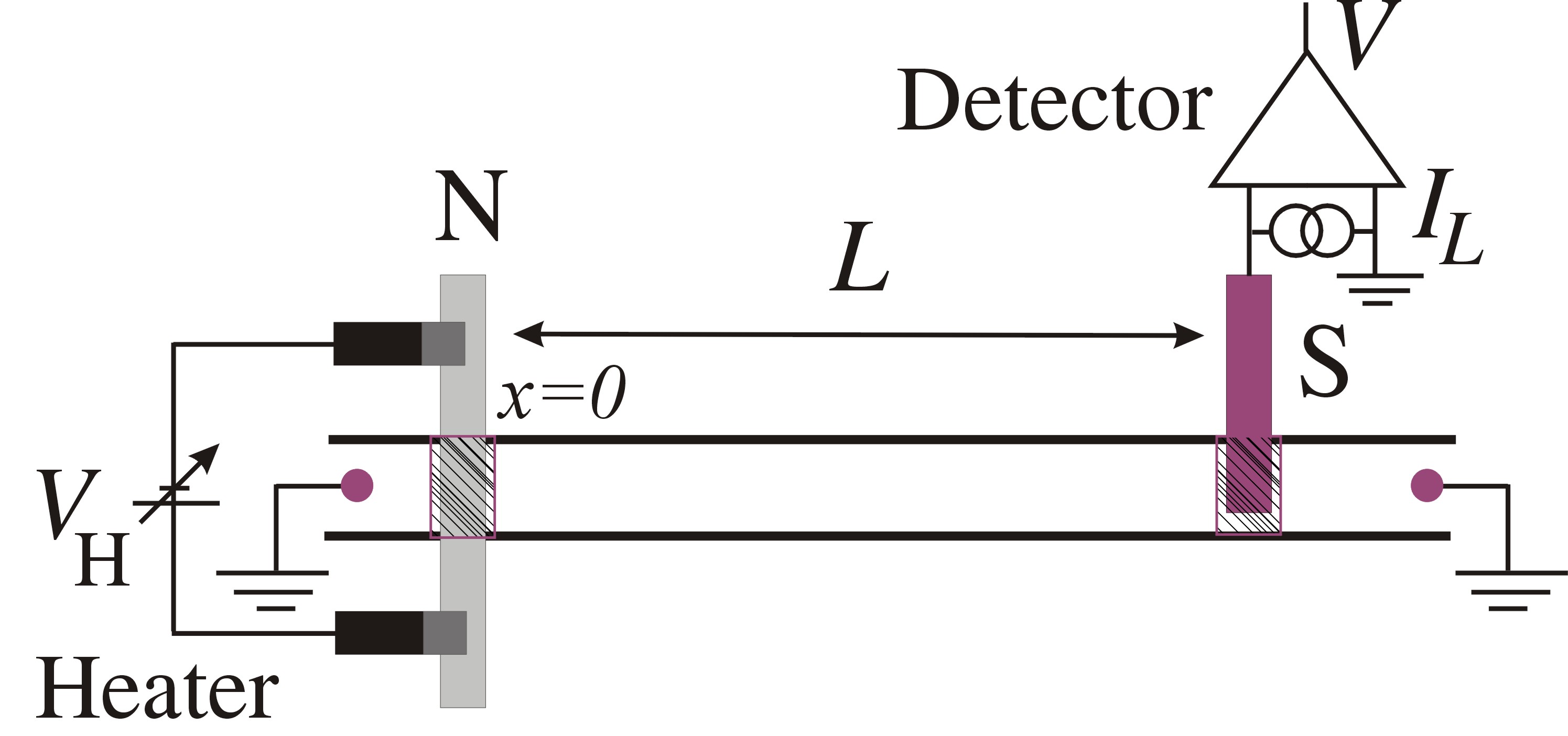}
\caption{Disordered wire of symmetry class D connected to a normal terminal $N$ ('source') and
a superconducting terminal $S$ ('drain') via two tunnel contacts. % with the conductances $g_{a,b} \ll 1$. 
The inflow of heat into the wire is generated in the $N$--terminal and is detected by the $S$--terminal
at a distance $L$ from the source. The $S$-terminal is biased by a voltage $V$. }
\label{fig:Wire_TT}
\end{figure}

For concreteness, we consider a  topological superconductor of class $\mathrm{D}$,
i.e.~a spin-rotation and time-reversal symmetry broken system supporting Majorana end
states in topological regimes. It may be driven across the topological phase transition by application of a  parallel magnetic filed~\cite{Lutchyn:2010,Oreg:2010, Albrecht:2016}. The conserved quantity
relevant to transport in a superconductor is \emph{heat}.  We imagine the heat current induced by
a heater, e.g. a normal conducting electrode, weakly coupled to the wire at
some point $x\equiv 0$, see Fig.~\ref{fig:Wire_TT}. We consider a protocol where at time $t=0$ the temperature
undergoes a sudden change $T\to T'$ such that the local distribution
function at $x=0$ changes as $\delta f(\epsilon,t)\equiv\Theta(t) \delta f(\epsilon)$, where
$\delta f(\epsilon)=f_{T'}-f_T$ and $f_T$ is the Fermi distribution function. The induced heat
current at a coordinate $x'\equiv L$ along the wire is best monitored by a \emph{superconducting} readout electrode which has the
convenient property~\cite{Giazotto:2006} that it converts heat into electrical current $I_L(t)$. Standard
linear response theory shows that to the first order in the temporal
Fourier transform $\delta f(\epsilon,\omega)={\delta f(\epsilon)}/{i \omega}$ of
the distribution at the heater electrode a current
\begin{equation}
\label{eq:j1_w}
 I_L(\omega) \propto (i \omega)^{-1}\int d\epsilon 
\Bigl[\nu(\epsilon  +  V ) - 
\nu(\epsilon - V ) \Bigr]  \Pi_L(\epsilon,\omega) \delta f(\epsilon),
\end{equation}    
is induced. Here $\nu$ is the BCS density of states in the probing terminal, assumed
to be biased by a voltage $V$ relative to the wire, and
$\Pi_L(\epsilon,\omega)=\left\langle G_{0,L}^+(\epsilon+\frac{\omega}{2})
G^-_{L,0}(\epsilon-\frac{\omega}{2}) \right\rangle$ is a response kernel describing particle
propagation inside the quantum wire in terms of its retarded and advanced Green
functions $G^{\pm}_{x,x'}(\epsilon)$.  Assuming a voltage $V\gtrsim \Delta$ exceeding the BCS
gap (so that the superconductor admits quasiparticle current) a linearization in
$\epsilon\sim T\ll V$ leads to a representation of the current $ I_L(\omega) \propto
(i \omega)^{-1}\nu'(V) \int d\epsilon
\,\epsilon\,  \Pi_L(\epsilon,\omega) \delta f(\epsilon)$ entirely in terms of the response kernel and the driving source. 

\begin{figure}[t]
\includegraphics[width=7.5cm]{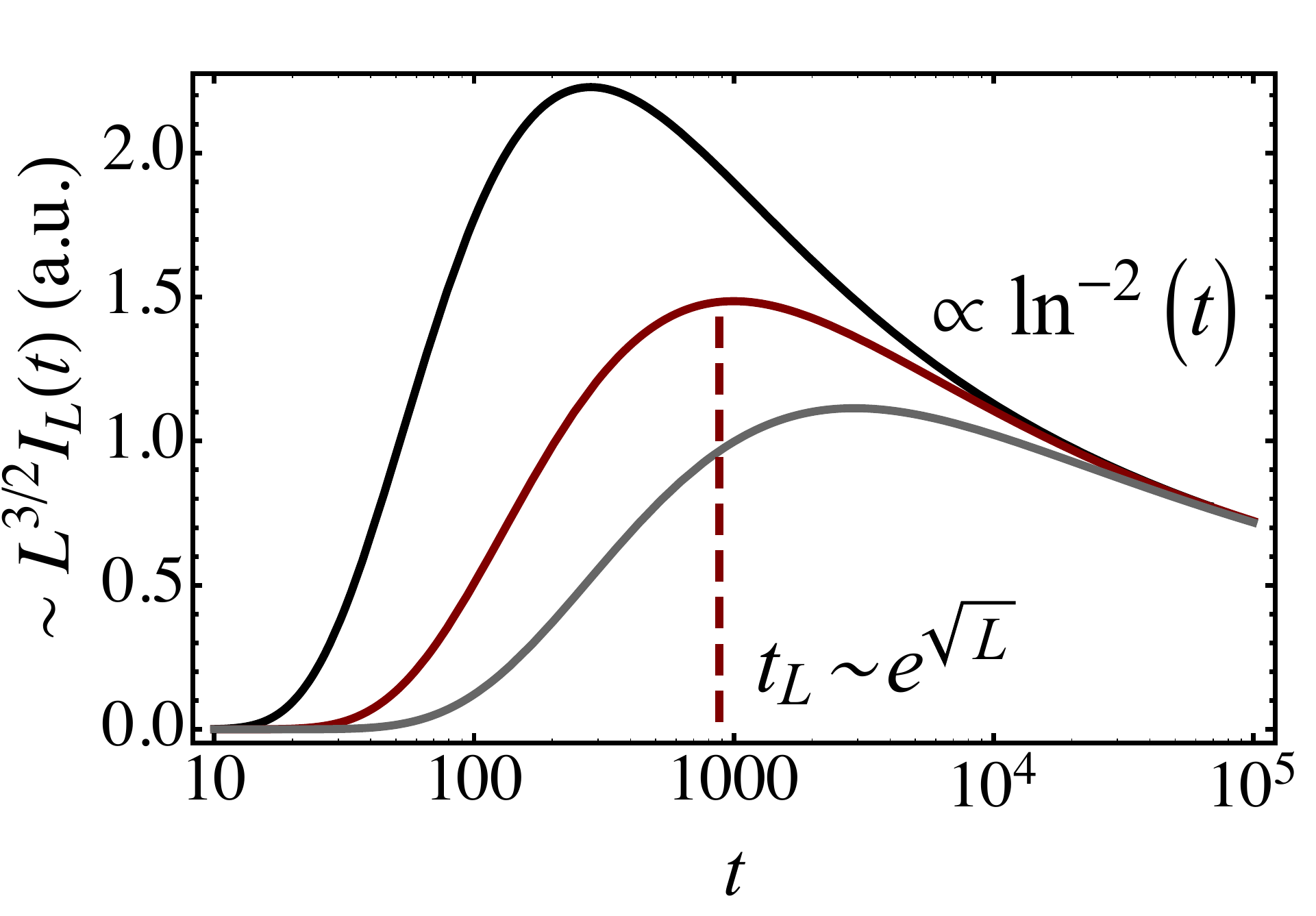}
\caption{
'Quantum' Sinai diffusion through the disordered quantum wire of class D at criticality 
as measured by the current $I_L(t)$ in the superconducting detector shown as the function of time $t$ 
for three separations $L/\xi_0=4,6,8$ between the heater and the detector.  
}
\label{fig:I_L}
\end{figure}

For generic energy arguments $\epsilon$ of the participating propagators the kernel
$\Pi_L$ is subject to  strong Anderson localization and  short-ranged on the length
scale $\xi_0$. Deviations from this behavior occur at a topological phase transitions
and at energies $\epsilon,\omega\to 0$ approaching the band center. 
For  $|\epsilon|\gtrsim \omega$ the kernel $\Pi_L(\epsilon,\omega)\simeq
\Pi_L(\epsilon,0)$ is suppressed due to the detuning of the Green function
energies off the band center. Our analysis detailed below predicts a spatial decay
like $\Pi_L(\epsilon,0)\sim \exp(-L/\xi_\epsilon)$ with an effective localization
length $\xi_\epsilon \propto \xi_0 \ln^2(\epsilon/\Delta_\xi)$, where $\Delta_\xi$ the
average single particle level spacing of a system of extension $\sim \xi_0$. 
This energy range contributes nearly instantaneous, but exponentially small 
response of the form  $I^>_L(t)\sim \Theta(t) \exp\{-L/\xi_{T'}\}$,  provided  
$T' < \Delta_{\xi}   e^{-(L/\xi_0)^{1/3}}$ and $I^>_L(t)\sim \Theta(t) \exp\{-(L/\xi_{0})^{1/3}\}$ 
in the opposite limit. 
Turning to the low
energy contribution  $|\epsilon|<\omega$, where  $\Pi_L(\epsilon,\omega)\simeq \Pi_L(0,\omega)\equiv
\Pi_L(\omega)$, the calculation discussed below yields 
\begin{equation}
                                                                    \label{eq:Pi_approx_D_k}
\Pi_L(\eta) \sim \frac{1}{\eta^2 \ln^5(1/\eta)} \sum_{n=1}^{\infty} n^2 e^{-n^2 L/\xi_\eta},
\quad \xi_\eta = \frac{\xi_0}{2\pi^2}\ln^2 \eta,
\end{equation}
where $\eta= - i \omega/\Delta_\xi$ is the dimensionless frequency.
Equation~\eqref{eq:Pi_approx_D_k} for the band center correlation function is the main
technical result of our work. This function replaces what would be a `diffusion propagator' in a conventional disordered metal. It is straightforward to check that Laplace transformation in $\eta$ leads to Sinai-diffusion $L\sim \ln^2(t)$ scaling.
To obtain the resulting current, we notice that  $ \int_{|\epsilon|<\omega} d \epsilon\, \epsilon\, \delta f(\epsilon) \sim
\omega^2$ (for 
$T<\omega<T'$) yields a factor to be interpreted as the diminishing support of states
contributing to the phase coherent transport at large times. Combining all factors and
performing a Laplace transformation $\eta\to t$ we finally obtain 
\begin{equation}
                  \label{eq:time-dependent}
I_L(t) \propto \frac{1}{\ln^5 (t\,\Delta_\xi)} \sum_{n=1}^{\infty} n^2 \exp\left\{-n^2\,\frac{2\pi^2 L}
{\xi_0 \ln^2 (t\,\Delta_\xi) }\right\} .
\end{equation}
Fig. \ref{fig:I_L} shows the current for several separations $L$. The signal reaches a maximum at times 
$t_L\!\!\sim\!\Delta_\xi^{-1} e^{\sqrt{L/\xi_0}}$ where it scales as $L^{-5/2}$. The temporally non-monotonous profile of the current results from the very slow buildup of the resonant conduction channel competing with the diminishing number of coherently contributing states at large times. In the rest of the paper we will discuss how the sub-diffusive form of the response kernel responsible for the unconventional thermal conduction properties of the system can be understood from first principles. 

We aim to explore transport in non-perturbative regimes and at finite frequencies
which makes supersymmetric field theory the method of choice. Though detailed
calculations leading to Eq.~\eqref{eq:Pi_approx_D_k} and presented in Supplemental Material~\cite{SM}   
are technically involved, the sketch below is meant to summarize
the  main ideas of the construction in concise terms. In supersymmetric field
theory, our correlation function  $\Pi_{x-x'}(\omega)\! = \langle Q_{x,12}^\mathrm{bf} Q_{x',21}^\mathrm{fb}\rangle$ is obtained as a  functional
average over $4\times 4$ supermatrix~\cite{Altland:2015} fields $Q^{\alpha \alpha'}_{\tau \tau'}$ carrying indices $\alpha=\mathrm{b,f}$ discriminating between commuting and anti-commuting entries, and `particle/hole' indices $\tau=1,2$. Conceptually, the functional integral is over the Goldstone mode manifold $Q\in \mathrm{SpO}(2|2)/\mathrm{U}(1|1)$ whose $(2\times
2)$-dimensional $\mathrm{ff}$-sector $\mathrm{O}(2)/\mathrm{U}(1)$
contains only two discrete elements, representable, e.g., as Pauli matrices
$\pm \tau_3$ in particle-hole space. This means that matrix-manifold is `disconnected', and that the
corresponding field theory must include the option of `kinks' between its two
connectivity components. Localization in this framework is a consequence of a
proliferation of those kinks. In this regard the freezing of jumps between
different field sectors and the proliferation of real-space jumps between sectors of alternating
topological index, taking place at the topological transition~\cite{Motrunich:2001}, are phenomena {\em dual} 
to each other. Kink formation is best described within the framework of a  `granular'
model comprising a chain of disordered superconducting quantum dots coupled by $2N$
quantum channels each characterized by a deterministic transmission $t_k$, where $k=1,\ldots 2N$.
The field theoretical action describing the system after disorder averaging is 
given by~\cite{Efetov-book,Nazarov-book,Kamenev-book}
\begin{equation}
\label{eq:Action_D}
	 \!\!S=\!\sum_{x,k} \frac{1}{4}\mathrm{str}\ln\!\left[1\!-\!\frac{ t_k^2}{4}\big(Q_x-Q_{x+1}\big)^2\right]-\frac{i\omega}{2\Delta_0}\,\mathrm{str}\left(Q_x \tau_3\right)\!,
\end{equation}        
where $x=1,\ldots,L$ labels the grains, $\tau_j$ are Pauli matrices in particle-hole space, and $\Delta_0$ is the granular level spacing.   
The bare \emph{fugacity}, $\chi_0$, corresponding to a kink between two granules $x\to x+1$  is given by the action (\ref{eq:Action_D}) evaluated on configurations whose fermion blocks $Q^{\mathrm{ff}}_x=\pm \tau_3$ and $Q^{\mathrm{ff}}_{x+1}=\mp\tau_3$ belong to opposite parts of the manifold. The substitution of these configurations leads to   
$
	\chi_0=e^{-S_\mathrm{kink}} = 
\prod_{k=1}^{2N} r_k  ,
$  
where $r_k=\pm (1-t_k^2)^{1/2}$ are the channel reflection coefficients. It was shown \cite{Fulga:2011} that on the non-topological side of
the transition all $r_k$ are positive, while on the topological side one (or odd
number) of $r_k$ must be negative. Thus the kink fugacity $\chi_0$ represents the
bare topological index. A critical configuration is reached when $\chi_0=0$,
i.e. when  at least one channel goes completely transparent, leading to vanishing kink fugacity.

The beauty of the one-dimensional theory is that it may be  solved exactly by
transfer-matrix methods. To this end we  introduce the distribution function
$\Psi(Q,x)=\int^{Q_x=Q}_{Q_{-\infty}=\tau_3}{\cal D}Q'\, e^{-S[Q']}$  and derive an evolution equation in the
form $\Psi(Q,x+1)=\Psi(Q,x)+\xi_0^{-1}\hat {\cal H} \Psi(Q,x)$, where $\xi_0\equiv g_0/2$ the bare localization length expressed in terms of the Landauer
inter-grain conductance $g_0=\sum_k t_k^2$, and  $\hat {\cal H}$ is a
second order differential operator acting on the variables parameterizing $Q$. Referring for an explicit expression of this Schr\"odinger like operator to~\cite{SM}, here we note two of its salient properties: due to the disconnected field integration
domain, $\Psi=(\Psi_+,\Psi_-)^T$, actually is a two component \emph{spinor} where $\Psi_\pm$ is
the probability to abide in the sector of $Q^{\mathrm{ff}}=\pm \tau_3$. Correspondingly, $\hat {\cal H}$  carries a   $2\times 2$ matrix structure whose off-diagonal elements generate kinks and thus are  proportional $\chi_0$. In the static case, $\omega=0$, the ensuing `discrete time' ($x$) spinor Schr\"odinger
equation can been solved analytically~\cite{Altland:2015} and describes
how both the average thermal DC conductance $g(L)$ and  average topological index $\chi(L)$ evolve with system size, Fig.~\ref{fig:flow}. In the following we focus on the quantum critical regime, $\chi(L)=0$, where the theory becomes ``spin'' conserving and a sub-Ohmic decay $g(L)\sim 1/\sqrt{L}$ signals delocalization.

Turning to the discussion of \emph{dynamical correlations}, $\omega\not=0$, we  notice that the first `kinetic' term of the path integral action Eq.~\eqref{eq:Action_D}
exhibits a high degree of rotational symmetry, which is partly broken by the
`potential' proportional to $\omega$. This structure motivates a parameterization of
the matrix fields in terms of three `angular degrees of freedom' (two of which are
anti-commuting), and one  `radial' variable $y$.  The formal analogy
between the transfer matrix equation and a time dependent Schr\"odinger equation 
suggests to consider the eigenfunctions $|\Psi_{n,l}\rangle$
diagonalizing now spin-conserving evolution operator as $\hat{\mathcal{H}} |\Psi_{n,l}\rangle
=\epsilon_{n,l}  |\Psi_{n,l}\rangle$ where $l$ and $n$ play the role of an  azimuthal and a radial quantum number, respectively, and $\epsilon_{n,l}$ are the corresponding eigenvalues. Much as in the solution of a quantum mechanical hydrogen problem, it turns out that the problem is separable and that the crucial radial part of the eigenfunctions $R_{n,l}(y)$ is governed by the one-dimensional equation   
\begin{equation}
\label{eq:D_Schroed_l}
\left[ -\frac{1}{2}\partial_y^2 + V_{\rm eff}(y) + \eta(\cosh2y-1)\!\right]\! R_{n,l}(y) = \epsilon_{n,l} R_{n,l}(y) ,
\end{equation} 
where the effective potential  
\begin{equation}
                                         \label{eq:Poschl-Teller}
V_{\rm eff}(y) = 
{\left( \frac 18 - \frac{l^2}{2}\right)}\frac{1}{\cosh^{2} y} + {\left( \frac 38 + \frac{l^2}{2}\right)}
\frac{1}{\sinh^{2} y}, 
\end{equation}
represents `centrifugal forces' in sectors of fixed angular momentum $l$. As in a rotationally symmetric quantum problem, the strong `central potential' $\sim \eta$ leads to confinement of the radial coordinate $y$ which in turn renders the spectrum discrete. Focusing on the case of interest $\eta\ll 1$ and matching the asymptotic solutions of the equation in two overlapping intervals 
$y\gg 1$ and $y\ll \frac 12\ln\eta^{-1}$, we obtain~\cite{SM}
\begin{equation}
                                                        \label{eq:spectrum}
 \epsilon_{n,l} = \frac{k_{n,l}^2}{2}, \quad\quad k_{n,l}= \frac{2\pi n}{\ln \eta^{-1}}; \quad\quad n=1,2,\ldots
\end{equation} 
which may be understood as the spectrum of a rectangular quantum well of width $y=\ln \eta^{-1}$. 
To extract physical information from this result, we represent the correlation function as a spectral decomposition $\Pi_{x-x'}(\omega)\! = \langle Q_{x,21}^\mathrm{bf} Q_{x',12}^\mathrm{fb}\rangle=\sum_{n} \Gamma_{n} \bar\Gamma_{n} \,\, e^{-\epsilon_{n,l} |x-x'|/\xi_0}$, where $\Gamma_{n}=\langle 0|Q^\mathrm{bf}_{21}|\Psi_{n,1} \rangle $ and we assumed a system of a large size $L_0\gg |x-x'|$ such that the evolution outside the observation interval $[x,x']$ is governed by the zero-energy ground state $|\Psi_{0,0}\rangle \equiv |0\rangle$. We also note that a single `coordinate function' $Q$ excites $|0\rangle$ only up to angular momentum $l=1$. Computing the matrix elements $\Gamma_{n}$ as detailed in~\cite{SM}, we obtain the final result~\eqref{eq:Pi_approx_D_k}.

Remarkably the correlation functions similar~\eqref{eq:Pi_approx_D_k} was obtained by Balents and Fisher~\cite{Balents:1997} as {\em single-particle } Green functions  of a single channel chain of the different symmetry class BDI.  This coincidence hints at a strong source of universality of the Sinai diffusion class which is not yet fully understood. We also note that the propagator of the classical
Sinai problem~\cite{Bouchaud:1990,Comtet:1998}, while similar to \eqref{eq:Pi_approx_D_k}, contains different matrix elements, resulting 
in a saturation of the response on $\sim \Theta(t)$ signal at a large time. This reflects the fact the classical problem does not rely on long time coherence.      

Summarizing, we have shown that disordered topological quantum wires compromise between the generic dominance of Anderson localization in low dimensions and the topologically enforced buildup of long range correlations at a quantum critical point by exhibiting an ultra-slow transport, a quantum generalization of the classical Sinai diffusion. While our discussion focused on the example of a multi-channel class D quantum wire, the phenomenon is universally present in all five one-dimensional symmetry classes. We discussed how Sinai diffusion leads to unconventional signatures in quantum transport, including the temporally non-monotonous propagation of thermal current pulses. The question whether similarly exotic phenomena are to be observed at topological quantum critical points in higher dimensions remains open.

%%%%%%%%%%%%%%%%%%%%%%%%%%%%%%%%%%%%%%%%%%%%%%%%%%%%%%%%%%%%%

%\noindent
{\emph{Acknowledgments:} 
We acknowledge discussions with I. Gruzberg and M. Westig. The research of A.K. was supported by NSF grant
DMR1306734.

%%%%%%%%%%%%%%%%%%%%%%%%%%%%%%%%%%%%%%%%%%%%%%%%%%%%%%%%%%%%%

\bibliography{Literature}	% *.bib-Datei
\bibliographystyle{apsrev-nourl}

%================Supplementary===============
\supplementarystart

\centerline{\bfseries\large SUPPLEMENTAL MATERIAL}
\vspace{6pt}

\centerline{\bfseries\large Sinai Diffusion at Quasi-1D Topological Phase Transitions}
\vspace{6pt}
\centerline{Dmitry~Bagrets, Alexander~Altland, and Alex~Kamenev}
\begin{quote}
%In this Supplemental Material we provide a number technical details related to the derivations of our results %presented in the main text of the paper. 
In this SM we derive and solve the transfer matrix equation and evaluate the propagator of Sinai diffusion.
\end{quote}

\maketitle

\subsection{Transfer matrix method}
In this section we discuss the transfer matrix Hamiltonian at the critical regime. The proliferation of kinks in this
case is suppressed and we can concentrate on the (+) sector of the Goldstone's manifold. Following our previous 
study (see Ref.~\cite{Altland:2015s}, section V and Appendix B), 
we parametrize it by two real ($y,\alpha$) and two Grassmann ($\bar\xi, \xi$) 
coordinates $z=(y,\alpha, \bar\xi, \xi)$ with the metric
$dg_+ = %- \frac 18 {\rm str}\left( dQ dQ\right) = 
g_{ij} dz^i dz^j = dy^2 + \sinh^2 2y \,d\alpha^2 + 2 \sinh^2 y\, d\bar\xi d\xi$.
With the Jacobian $J_+ = \sqrt{{\rm sdet} g_+}=2 \coth y$, the former defines the transfer matrix Hamiltonian
${\cal H} = -\frac{1}{2} J_+^{-1}  \partial_i (  g^{ij} J_+ \partial_j ) + V_\eta(y)$, %\qquad 
where $V_\eta = \eta (\cosh 2y - 1)$ is the potential energy resulting from the frequency term of the 
action~(\ref{eq:Action_D}) and $\eta = -i\omega/\Delta_\xi$. In what follows we take 
$\eta \in \mathbb{R}^+$ thereby changing from the Fourier to Laplace transform.

To find the eigenfunctions $\Psi_k(z)$ and the spectrum $E_k$ of ${\cal H}$ --- here $k=\{n,l,\bar\sigma,\sigma\}$ 
is the set of quantum numbers defined below --- we perform the Sutherland transformation, 
$\Phi_k(z) = \sqrt{J_+(y)}\Psi_k(z)$, which brings the
transfer matrix Hamiltonian to the form ${\cal H} = {J_+}^{1/2} { H} {J_+}^{-1/2}$ with
\begin{equation}
H = - \frac{1}{2}\partial_y^2 + V_{\rm PT}(y) + V_\eta(y) - \frac 12 \sinh^{-2}(2y)\, \partial_\alpha^2 -
\sinh^{-2}(y)\,\partial_{\bar\xi}\partial_\xi, \qquad
V_{\rm PT}(y) = \frac 18 \cosh^{-2} y + \frac 38 \sinh^{-2} y. 
\end{equation} 
Here $V_{\rm PT}(y)$ is known as the 'Poschl-Teller' potential.
The zero energy state $|0\rangle\equiv \Phi_0(y)$ of 
the Hamiltonian $H$ is 'spherically' symmetric and at $y \gtrsim 1$ approximately reads as 
$\Phi_0 (y) \simeq   - 2 K_0(\sqrt\eta e^y)/\ln \eta$, cf. Ref.~\cite{Altland:2001s}. 
Its normalization is the consequence of $\Phi_0(0)=1$ guaranteed by the supersymmetry.
Considering further excited states $| k \rangle \equiv \Phi_k(z)$, 
we separate angular and radial variables and represent the wave function as 
$\Phi_k(z) = {\cal R}_{k}(y)\times e^{2 i l \alpha} \, e^{\bar\xi \sigma + \xi \bar\sigma}$. Here
$l\in\Bbb{Z}$ and Grassmanns ($\sigma, \bar\sigma$) are angular quantum numbers while ${\cal R}_{k}(y)$
satisfies to the radial Schr\"odinger equation 
\begin{equation}
\label{eq:Schr_R}
\Bigl( -\frac{1}{2}\partial_y^2 + V_{\rm eff}(y) + V_\eta(y) + 
\sinh^{-2}(y)\,{\bar \sigma \sigma }\Bigr) {\cal R}_k(y) = E_k {\cal R}_k(y),
\end{equation}
where $V_{\rm eff}(y)$ is the effective potential~(\ref{eq:Poschl-Teller}). 
We note that $\bar\sigma \sigma$ is the nilpotent element of the Grassmann algebra. 
Therefore the spectrum and eigenstates of (\ref{eq:Schr_R}) must have the form  
$E_k = \epsilon_{n,l} + \bar\sigma \sigma\, \epsilon_{n,l}' $ and 
${\cal R}_k(y) = R_{n,l}(y) + \bar\sigma \sigma R_{n,l}'(y)$, resp., 
where $n=1,2,\dots$ is the radial quantum number.
As shown below in Sec.\ref{sec:ME}, 
only the 0th--order terms are needed to find the propagator $\Pi_L(\eta)$ of quantum Sinai diffusion. 
We thus construct the asymptotic form of the radial wave function
$R_{n,l}(y)$ in the limit  $\eta \ll 1$ in the next section and then evaluate $\Pi_L(\eta)$ in Sec.\ref{sec:ME}.    

\subsection{Radial wave function}
\label{sec:Radial_R}

To solve the one-dimensional equation~(\ref{eq:D_Schroed_l}) for the radial part $R_{n,l}(y)$ 
we introduce momenta $k_{n,l}$, such that energies $\epsilon_{n,l} = \frac 12 k_{n,l}^2$ and
split $y$-axis in three intervals: (I) 'small' angles, $0 < y < 1 $; (II) 'intermediate' ones, 
$ 1 < y < \frac 12 \ln (1/{\eta})$ and (III) 'large' angles, where $y > (1/2 )\ln (1/{\eta})$.

In the intervals II \& III one can approximate Eq.~(\ref{eq:D_Schroed_l}) by 
$\left[ -\partial_y^2 +  \eta e^{2y}\right] R_{n,l}(y) = k_{n,l}^2 R_{n,l}(y)$.
Up to a normalization factor which is found below, the solution of this equation is a modified  
Bessel function $R_{n,l}(y) \propto K_{ik_{n,l}}(\sqrt{\eta} e^y)$ taken at imaginary index. 
On taking $K_\nu(z)$ at small argument, $R_{n,l}(y)$ in the interval II 
is reduced to the plane wave
\begin{equation}
\label{eq:R_n_region_II_A}
R_{n,l}(y) \propto A(k_{n,l}) e^{i  k_{n,l} y } + A^*(k_{n,l}) e^{-i k_{n,l} y }, 
\qquad A(k) = \Gamma(-ik) \left(\eta/ 2\right)^{ik/2}.
\end{equation}
On the other hand, if $y$ is restricted to the intervals I \& II, one can equally well neglect by $\eta$-dependent
part of the potential and solve the equation
$\left[ -\partial_y^2 +  V_{\rm eff}(y)\right] R_{n,l}(y) = k_{n,l}^2 R_{n,l}(y)$.
It admits the exact solution in the form  
$R_{n,l}(y) = (1-u)^{-ik_{n,l}/2} u^{\alpha_l/2} g(u)$, where we've defined a new variable $u=\tanh^2 y$.
In this ansatz the exponent $\alpha_l = \frac12 + \sqrt{1+l^2}$ and  
$g(u) = {}_2F_1(a,b,c; u)$ is the hypergeometric function with parameters
\begin{equation}
a = \left(1 - l + \sqrt{1+l^2} - ik\right)/2,
\quad
b = a + l, \quad
c= 1 + \sqrt{1+l^2}.
\end{equation}
In the interval II such radial wave function has the asymptotic expansion 
\begin{equation}
\label{eq:R_n_region_II_B}
R_{n,l}(y) \propto B^*(k_{n,l}) e^{i k_{n,l} y } + B(k_{n,l}) e^{-i k_{n,l} y }, 
\qquad B(k) = {\Gamma(-ik)}/\bigl({\Gamma(a) \Gamma(b)}\bigr).
\end{equation}
The plane waves~(\ref{eq:R_n_region_II_A}) and (\ref{eq:R_n_region_II_B}) should match in the interval II
which is only possible if momentum $k_{n,l}$ is quantized. 
On introducing the scattering matrices and phase shifts from the left \& right potential barriers,
$S_L(k) = {B(-k)}/{B(k)} = e^{-i\phi_L(k)}$~and~$S_R(k) = {A(-k)}/{A(k)} = e^{-i\phi_R(k)}$,
one arrives at the quantization condition $\phi_R(k_{n,l}) + \phi_L(k_{n,l}) = 2\pi n$.
For small momenta $k_{n,l} \ll 1$ and with log-accuracy it is simplified to 
$\eta^{-ik_{n,l}} \simeq e^{2i\pi n}$. This leads to the quantized momenta and energies $\epsilon_{n,l}$
as stated in the main text, see~Eq.~(\ref{eq:spectrum}).

To find the properly normalized radial wave function, we note that the main contribution to the integral
$\int_0^{+\infty } R^2_{n,l}(y) dy=1$ comes from the interval II.
Using the scattering shift, the wave function in this region (when extrapolated from the interval III) becomes
$R_{n,l}(y) \propto |A(k_{n,l})| \cos\left( k_{n,l} y +  \frac 12 \varphi_R(k_{n,l})\right)$.
On the other hand, the same wave function when found within the semiclassical approximation should read
$R_{n,l}(y) = ({C_{n,l}}/{\sqrt{k_{n,l}}}) \cos\left( k_{n,l} y + \phi\right)$, 
where the normalization constant is fixed by 
$C_{n,l}^2 = ({k_{n,l}}/{\pi})({\partial k_{n,l}}/{\partial n})$. 
On comparing these two representations 
we conclude that if the angle $y$ is not 'small' the normalized radial wave function reads
\begin{equation}
\label{eq:Rn_y_II_III}
R_{n,l}(y) = \left(\frac{1}{\pi} \frac{\partial k_{n,l}}{\partial n} \right)^{1/2} 
|A(k_{n,l})|^{-1} K_{ik_{n,l}}(\sqrt{\eta} e^y),
\qquad |A(k)|^{-1} = \left(\frac{k \sinh k\pi}{\pi}\right)^{1/2}, \qquad y \gtrsim 1.
\end{equation}
We use this important intermediate result below to find the series expansion of the kernel $\Pi_L(\eta)$. 

\subsection{Matrix elements}
\label{sec:ME}

The response kernel at zero energy
($\epsilon = 0$) is related to a correlation function of the field theory 
via the identity   
\begin{equation}
\label{eq:Pi_def}
\Pi_L(\omega)=\sum_{\alpha\beta}\left\langle G_{0,L}^{+,\alpha\beta}({\omega}/{2}) G^{-,\beta\alpha}_{L,0}(-{\omega}/{2}) \right\rangle_{\rm dis} = -
\sum_{\alpha\beta}\left\langle G_{0,L}^{+,\alpha\beta}({\omega}/{2}) G^{+,\alpha\beta}_{0,L}({\omega}/{2}) \right\rangle_{\rm dis}
= -
\langle  Q_{21}^\mathrm{fb}(0) Q_{12}^\mathrm{bf}(L) \rangle.
\end{equation}
Here Greek indices account for the transverse quantum channels in the wire and 
Latin indices (1,2) refer to the ph-space within the bf- and fb-blocks of $Q$ matrix.
In the chosen parametrization these matrix elements have the form
$Q^{\rm bf}_{21} = - e^{-2 i\alpha} \xi \sinh 2y$ and $Q^{\rm fb}_{12} = (Q^{\rm bf}_{21})^*$.
Employing further a spectral decomposition, we write
\begin{equation}
\label{eq:cal_Ak}
\Pi_L(\omega)  = -\sum_{n,l}\int d\bar\sigma d\sigma\, 
{\Gamma}_k \bar {\Gamma}_k e^{- E_k L/\xi_0 },
\qquad
{\Gamma}_k = \langle 0 | Q_{21}^{\rm bf} | k \rangle = 
\int\limits_0^{+\infty} dy \int\limits_0^{2\pi} d \alpha \int d\bar\xi d\xi\, \Phi_0(y) Q_{21}^{\rm bf}   \Phi_k(z)
\end{equation} 
where $\Gamma_k$ is the matrix element and the analogous expression holds for $\bar {\Gamma}_k$. 
Observe that fields $Q_{21}^{\rm bf}$ and $Q_{12}^{\rm bf}$ are linear in Grassmanns $\xi$ and $\bar\xi$, resp. 
Using the explicit form of $\Phi_k(z)$ and performing the integral over Grassmanns in Eq.~(\ref{eq:cal_Ak})
it is then straightforward to see that the nilpotent part
$\sim R_{n,l}'(y)$ of the radial wave function does not contribute to the matrix elements. They are
simplified to ${\Gamma}_k = - \sigma \Gamma_{n,l}$ and $\bar {\Gamma}_k = \bar \sigma \Gamma_{n,l}$, with
$\Gamma_{n,1} = \int_0^{+\infty} \Phi_0(y) R_{n,1}(y) \sinh (2y)  dy$
being non-zero only for $l = 1$.
The leading contribution to the latter integral comes from the region $ y \gtrsim 1$, thus we can use the 
asymptotic form~(\ref{eq:Rn_y_II_III}) for $R_{n,1}(y)$.
On approximating $\sinh(2y) \simeq \frac 12 e^{2y}$, changing the variable of integration to
$z= \sqrt\eta e^y$ and using the table integral 
$\int_0^{+\infty} z K_0(z) K_{i k}(z) = (k^2\pi^2/8)\sinh^{-2}(\pi k/2)$
we find for the matrix elements 
\begin{equation}
\label{eq:Mn}
M_n=\Gamma_{n,1}^2 \sim  \frac{k_{n,1}^2}{\eta^2\ln^2\eta} 
\left(\frac{\partial k_{n,1}}{\partial n}\right)\times \frac{k_{n,1}^3 \cosh (\pi k_{n,1}/2)}
{\sinh^3 (\pi k_{n,1}/2)} \,\, \overset{k_{n,1} \ll 1}{\longrightarrow } \,\,\frac{n^2}{\eta^2 \ln^5(1/\eta)}.
\end{equation}
Then after analytical continuation to imaginary frequency $\eta = - i\omega/\Delta_\xi$, the polarization
operator evaluates to 
\begin{equation}
\label{eq:Pi_series}
\Pi_L(\eta)  = -\sum_{n,l=1}\int d\bar\sigma d\sigma\, 
{\Gamma}_k \bar {\Gamma}_k e^{- E_k L/\xi_0 } = \sum_{n} M_n e^{- \epsilon_{n,1} L/\xi_0}.
\end{equation}
It is worth mentioning that $\bar \sigma \sigma \epsilon_{n,1}$--like correction 
to the spectrum does not contribute to this result. In the case of interest, $L \gtrsim \xi_0$,
the only essential momenta are small ($k_{n,1} \ll 1$) and the sum 
is simplified to our final result~(\ref{eq:Pi_approx_D_k}).

\end{document}